\documentclass[twocolumn,showpacs,preprintnumbers,amsmath,amssymb]{revtex4}

\usepackage{graphicx}
\usepackage{dcolumn}
\usepackage{bm}

\begin{document}


\title{Heating of quasiparticles driven by oscillations of the order parameter in short superconducting microbridges}

\author{D.Y. Vodolazov$^1$}
\email{vodolazov@ipm.sci-nnov.ru}
\author{F.M. Peeters$^2$,}
\affiliation{$^1$ Institute for Physics of Microstructures,
Russian Academy of Sciences, 603950,
Nizhny Novgorod, GSP-105, Russia \\
$^2$Departement Fysica, Universiteit Antwerpen, Groenenborgerlaan
171, B-2020 Antwerpen, Belgium}

\date{\today}


\begin{abstract}
We predict 'heating' of quasiparticles driven by order parameter
oscillations in the resistive state of short superconducting
microbridges. The finite relaxation time of the magnitude of the
order parameter $|\Delta|$ and the dependence of the spectral
functions both on $|\Delta|$ and the supervelocity $Q$ are the
origin of this effect. Our result is opposite to those of
Aslamazov and Larkin (Zh. Eks. Teor. Fiz. {\bf 70}, 1340 (1976))
and Schmid, Sch\"{o}n and Tinkham (Phys. Rev. B {\bf 21} 5076
(1980)) where 'cooling' of quasiparticles was found.
\end{abstract}

\maketitle

\section{Introduction}

We study theoretically the resistive state in short
superconducting microbridges (with length L less than the
coherence length $\xi(T)$) at temperatures close to the critical
one. This subject was widely discussed and studied in the 70-ies
of last century (for review see \cite{Likharev}) and has regained
renewed interest recently (see for example Refs.
\cite{Wang,Hazra}). Interesting phenomena occurring in such a
system, and which are still not completely understood, are: i) the
hysteresis of current-voltage (IV) characteristics at relatively
low temperatures \cite{Likharev} and ii) the 'foot'-like
(sometimes also called 'shoulder'-like - see Ref.
\cite{Aslamazov0}) feature in the IV characteristics observed
experimentally mainly in tin microbridges at low voltages and at
temperatures close to $T_c$ \cite{Guthmann,Gubankov,Octavio} (see
also \cite{Likharev}). The hysteresis is usually explained by
Joule heating while for the 'foot'-like structure several theories
were proposed \cite{Aslamazov,Golub,Schmid} that are based on the
idea that the quasiparticle distribution function is out of
equilibrium (overcooled) as a consequence of the variation in time
of the magnitude of the superconducting order parameter
$\Delta=|\Delta|e^{i\phi}$ in the superconducting microbridge. The
motivation for those theories comes from the fact that the energy
of the quasiparticles depends on $|\Delta|$ and when the
characteristic time scale for the variation of $|\Delta|$ is
smaller than the inelastic relaxation time $\tau_{in}$ of the
quasiparticle distribution function $f(\epsilon)$ the occupancy of
the states with energy $\epsilon$ may differ from the equilibrium
one (for a detailed discussion see the book of Tinkham
\cite{Tinkham}).

To simplify the analytical treatment of the problem the authors of
Refs. \cite{Aslamazov,Schmid} assumed that in the dynamic
(resistive) state $|\Delta|$ varies as fast as $\delta \phi$ and
to find the coordinate and time dependence of $|\Delta|(x,t)$ they
solved the {\it stationary} Ginzburg-Landau equation with a time
dependent $\delta \phi(t)$. Furthermore they assumed that: 1)the
relaxation term in the kinetic equations can be neglected when the
period of oscillations of the order parameter  is much smaller
than $\tau_{in}$, 2) the spectral functions depend only on the
local magnitude of the order parameter $|\Delta|$ (so called local
approach). We argue that for a realistic inelastic relaxation time
$\tau_{in}$, and even for a short microbridge $L \lesssim \xi(T)$
the nonequilibrium contributions to $f(\epsilon)$ strongly affects
the dynamics of $|\Delta|$ and results in a larger time scale for
$|\Delta|$ than for $\delta \phi$. Moreover the relaxation term in
the kinetic equations plays a very important role at any voltage
and therefore cannot be omitted. Taken together with the
dependence of the spectral functions on both $|\Delta|$ and $Q$
they provide an averaged 'heating' of the quasiparticles, instead
of 'cooling' \cite{Aslamazov,Schmid}. Some kind of 'cooling' of
quasiparticles at high voltages can be found only when one takes
into account additional terms in the kinetic equations which
couple the longitudinal $f_L$ (odd in energy) and transverse $f_T$
(even in energy) parts of
$2f(\epsilon)=(1-f_L(\epsilon)-f_T(\epsilon))$ due to the finite
spectral supercurrent (in previous works \cite{Aslamazov,Schmid}
these terms were omitted). This 'cooling' is not effective at low
voltages and 'heating' together with the large time scale of the
variation of $|\Delta|$ result in a hysteresis of the IV
characteristics for relatively large $\tau_{in}$. We should stress
that the time averaged 'heating' of quasiparticles is driven by
oscillations of $|\Delta|$ in the superconducting microbridge and
{\it not} by Joule dissipation $\sim {\bf j \cdot E}$ ({\bf j} is
the current density and {\bf E} is the electric field).

The paper is organized as follows. In section II we discuss the
theoretical model. In section III we present and discuss our
results. In section IV we present our conclusions and discuss the
possible origin of the experimentally found
\cite{Guthmann,Gubankov,Octavio} 'foot'-like structure in the IV
characteristics.

\section{Model}

To simulate the resistive state in a short superconducting
microbridge we use the kinetic equations derived in Refs.
\cite{Schmid2,Larkin,Kramer,Watts-Tobin} for 'dirty'
superconductors near $T_c$
\begin{widetext}
\begin{subequations}
\begin{eqnarray}
N_1\frac{\partial \delta f_L }{\partial t
}=D\nabla((N_1^2-R_2^2)\nabla \delta f_L)+D\nabla (j_{\epsilon}
f_T) -\frac{N_1}{\tau_{in}}\delta f_L-R_2\frac{\partial
f_L^0}{\partial \epsilon}\frac{\partial |\Delta|}{\partial t},
\end{eqnarray}
\begin{eqnarray}
\frac{\partial}{\partial t}N_1(f_T+e\varphi\frac{\partial
f_L^0}{\partial \epsilon})=D\nabla((N_1^2+N_2^2)\nabla f_T)+D
\nabla(j_{\epsilon}\delta f_L)- \frac{N_1}{\tau_{in}}
\left(f_T+e\varphi\frac{\partial f_L^0}{\partial
\epsilon}\right)-N_2|\Delta|\left(2f_T-\frac{\partial
f_L^0}{\partial \epsilon}\frac{\partial \phi}{\partial t}\right),
\end{eqnarray}
\end{subequations}
\end{widetext}
here $Q=(\partial \phi/\partial x-2eA/c)$ is a quantity which is
proportional to the superfluid velocity ($v_s=\hbar Q/m$),
$\varphi$ is an electrostatic potential, $\delta f_L=f_L-f_L^0$
and $f_L^0(\epsilon)=\tanh(\epsilon/2k_BT)$. $N_1$, $N_2$, $R_2$
are the spectral functions which should be found from the Usadel
equation for the normal $\alpha (\epsilon)=\cos
\Theta=N_1(\epsilon)+iR_1(\epsilon)$ and anomalous $\beta_1 =\beta
e^{i\phi}$, $\beta_2=\beta e^{-i\phi}$ ($\beta (\epsilon)=\sin
\Theta=N_2(\epsilon)+iR_2(\epsilon)$) Green functions
\begin{equation}
\hbar
D\frac{d^2\Theta}{dx^2}+((2i\epsilon-\frac{\hbar}{\tau_{in}})-\hbar
DQ^2 \cos\Theta)\sin\Theta+2|\Delta|\cos\Theta=0,
\end{equation}

Eqs. 1(a,b) are coupled through the finite spectral supercurrent
\cite{Larkin,Keizer} $j_{\epsilon}=Re(\beta_1\nabla
\beta_2-\beta_2 \nabla \beta_1 )/2=2N_2R_2Q$. Below we show that
the coupling terms in Eqs. 1(a,b) strongly influence $\delta f_L$
and the dynamics of the order parameter which is described by the
modified time-dependent Ginzburg-Landau equation
\begin{eqnarray}
\frac{\pi\hbar}{8k_BT_c}\frac{\partial \Delta}{\partial
t}+(\Phi_1+i\Phi_2)\Delta=\xi_{GL}^2\frac{\partial^2\Delta}{\partial
x^2}+\left(1-\frac{T}{T_c}-\frac{|\Delta|^2}{\Delta_{GL}^2}\right)\Delta
\end{eqnarray}

where $\xi_{GL}^2=\pi\hbar D/8k_BT_c$ and
$\Delta_{GL}^2=8\pi^2(k_BT_c)^2/7\zeta(3)$ are the zero
temperature Ginzburg-Landau coherence length and the corresponding
order parameter. Nonequilibrium parts of the quasiparticle
distribution function enters Eq. (3) via the potentials
$\Phi_1=-\int_0^{\infty}R_2 \delta f_Ld\epsilon/|\Delta|$ and
$\Phi_2=-\int_0^{\infty}N_2f_T d\epsilon/|\Delta|$. When $\delta
f_L$ is negative the potential $\Phi_1$ is positive and vice
versus. In some respect, from the structure of Eq. (3) it follows,
that one may introduce an 'effective' temperature for the
quasiparticles $T_{eff}(x,t)=T+\Phi_1(x,t)T_c$ and thus the
positive/negative sign of $\Phi_1(x,t)$ means local
'heating'/'cooling' of quasiparticles. We should stress that we
use the term 'effective' temperature only in order to give a
simple physical interpretation of our numerical results and
describe the {\it integral} effect of the nonequilibrium
distribution $f(\epsilon)$ which enters the equation for the order
parameter via the potentials $\Phi_1(x,t)$ and $\Phi_2(x,t)$.
Please note that the resulting nonequilibrium $f(\epsilon)$ cannot
be viewed as a Fermi-Dirac function with effective temperature
$T_{eff}(x,t)$ and electrostatic potential $\varphi(x,t)$.

The current and the electrostatic potential in the sample can be
found using the following equations
\begin{equation} 
j=\frac{\sigma_n}{e}  \left (\frac{|\Delta|^2Q}{4k_BT_c}+
\int_0^{\infty}\left( \left(N_1^2+N_2^2 \right)\nabla f_T
+j_{\epsilon} \delta f_L \right) d\epsilon  \right),
\end{equation}

\begin{equation} 
e\varphi=-\int_0^{\infty}N_1f_T d\epsilon / \int_0^{\infty}N_1
\frac{\partial f_L^0}{\partial \epsilon} d\epsilon,
\end{equation}

where $\sigma_n$ is the normal state conductivity. In metals we
have for the charge density $\rho \simeq 0$ and the condition $div
j =0$ is satisfied due to Eqs. (1b, 3, 5).

In the derivation of Eqs. 1(a,b) it was assumed that deviations
from equilibrium are small $\delta f_L,f_T \ll f_L^0$. It allowed
one to linearize the collision integral due to electron-phonon
collisions and to write it in the relaxation time approximation.
Furthermore, it was assumed that the inelastic relaxation time due
to electron-electron interactions is much larger than due to
electron-phonon and hence one can neglect the corresponding
collision integral.
\begin{figure}[hbtp]
\includegraphics[width=0.45\textwidth]{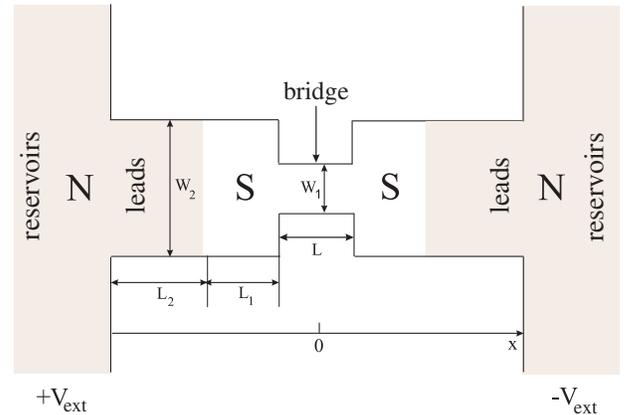}
\caption{Schematic illustration of the model system with the
geometric parameters.}
\end{figure}

To model the superconducting microbridge connected to
superconducting bulk leads in the current carrying regime we
consider the system shown in Fig. 1. It consists of normal bulk
reservoirs where the external voltage $V_{ext}$ is applied and the
constriction is modelled as a film with variable cross-section
$S_2 \gg S_1$ (here $S_{1(2)}=W_{1(2)}d_{1(2)}$ is the
cross-section, $W_{1(2)}$ is the width and $d_{1,2}$ is the
thickness of the film (in this way we model variable thickness
microbridge). The narrowest part of the film is called the
microbridge and the wide parts are assigned as being the leads.
The wide part of the film of length $2L_2$ is in the normal state
and the rest of the film is in the superconducting state (for
relatively small $V_{ext}$). In our calculations we chose $L_2/L_1
\gg 1$ and the length $L_1$ was taken large enough to neglect
nonequilibrium effects from the NS boundaries located at $x=\pm
(L_1+L/2)$ on the transport properties of the microbridge. The
large normal part of the film is needed because in the boundary
conditions for $f_L$ and $f_T$ enters the voltage $V$ (see for
example Ref. \cite{Keizer}) and almost all the voltage drop occurs
in the normal part of the film, effectively leading to a state
with applied constant current even when the microbridge transits
to the resistive state. Due to the large ratio $S_2/S_1 \gg 1 $
the transport current does not destroy superconductivity in the
wide part of the film and we mainly study the resistive state in
the microbridge. In our calculations we assume that the lateral
size of the film is much smaller than the London penetration depth
$\lambda$ (or effective penetration depth
$\lambda_{\bot}=\lambda^2/d_{1(2)}$) and therefore we may neglect
screening effects. Furthermore, we suppose that the current
density is distributed uniformly over the width of the film. This
assumption is strictly speaking not valid in the region where the
current transits from the wide part to the narrow one (and vice
versus). But because the current density is much smaller in the
wide part than in the narrow part we may neglect such type of
effects (at least for variable thickness microbridges) and
consider variables which are averaged over the width and thickness
of the film. Because the length of the microbridge is much larger
than its width we may assume an uniform current distribution over
the width of the microbridge. Thus our system is quasi 1D and we
only should take into account the continuity of $\Delta$,
$\Theta$, $f_T$, $\delta f_L$ and conservation of the
superconducting (Eq. 6(a) below), normal (Eq. 6(d)), energy (Eq.
6(c)) and spectral (Eq. 6(b)) currents at the points $x=\pm L/2$
through the boundary conditions
\begin{widetext}
\begin{subequations}
\begin{eqnarray}
S_2\partial \Delta/\partial x|_{\pm L/2}=S_1\partial
\Delta/\partial x|_{\pm L/2}
\\
S_2\partial \Theta/\partial x|_{\pm L/2}=S_1\partial
\Theta/\partial x|_{\pm L/2}
\\
S_2D ((N_1^2-R_2^2)\partial \delta f_L/\partial x+j_{\epsilon} f_T
)|_{\pm L/2}=S_1D (N_1^2-R_2^2) \partial \delta f_L/\partial x +
j_{\epsilon} f_T)|_{\pm L/2}
\\
S_2D ((N_1^2+N_2^2)\partial f_T/\partial x+ j_{\epsilon}\delta
f_L)|_{\pm L/2}=S_1D ((N_1^2+R_2^2)\partial f_T/\partial
x+j_{\epsilon} \delta f_L)|_{\pm L/2}
\end{eqnarray}
\end{subequations}
\end{widetext}

It is easy to find the solution to Eqs. (1-5) in the normal region
and therefore we only need to solve them in the superconducting
region (in the interval $-L_1-L/2<x<L_1+L/2$) with the following
boundary conditions: $\Delta|_{\pm (L_1+L/2)}=0$, $\Theta|_{\pm
(L_1+L/2)}=0$, $\delta f_{L}|_{\pm (L_1+L/2)}=0$, $f_{T}|_{\pm(
L_1+L/2)}=-e\varphi_{\pm} \cdot
\partial f_L^0/\partial E$, where $\varphi_{\pm} =\mp
V_{ext}\pm j_{\pm}L_2/\sigma_n $ and the current density $j_{\pm}$
in the points $x=\pm (L_1+L/2)$ could be found from Eq. (4).
\begin{figure}[hbtp]
\includegraphics[width=0.45\textwidth]{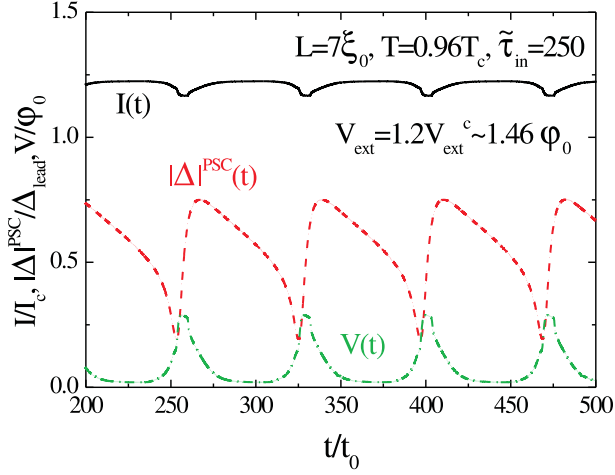}
\caption{(color online) Time dependence of the current, amplitude
of the order parameter in the center of the microbridge and
voltage drop over the microbridge at fixed external voltage
$V_{ext}$.  The results were found from numerical solution of Eqs.
(1-5) in the presence of coupling terms in Eqs. 1(a,b).  The
geometrical parameters of the system are: $W_2/W_1=10$, $L_2=500
\xi_0$, $L_1=22\xi_0$ and $L=7\xi_0$.}
\end{figure}

In our numerical calculations we use dimensionless units. The
order parameter is scaled by $\Delta_0$ ($\Delta_0=1.76
k_BT_c\simeq 0.57 \Delta_{GL}$ is the zero temperature order
parameter value in the weak coupling limit), distance is in units
of the zero temperature coherence length $\xi_0=\sqrt{\hbar
D/\Delta_0} \simeq 1.2\xi_{GL}$, time in units of
$t_0=\hbar/\Delta_0$ and temperature in units of the critical
temperature $T_c$. The current is scaled in units of
$j_0=\Delta_0\sigma_n/(\xi_0e)$, the superfluid velocity in units
of $Q_0=\hbar c/2e\xi_0$ and the electrostatic potential is in
units of $\varphi_0=\Delta_0/e$. It is useful to introduce the
dimensionless inelastic relaxation time $\tilde {\tau}_{in}=
\tau_{in}/t_0$ which is the main control parameter in the model
described by Eqs. 1(a,b).

We used the implicit method for the numerical solution of Eqs.
(1-3). The coordinate step of the grid was equal to $\xi_0$ (which
is much smaller than $\xi(T)$ for the considered temperature
interval $0.92<T/T_c<0.99$) and the time step varied from 0.5
$t_0$ up to 2 $t_0$ depending on the temperature. In our numerical
procedure we use an even number of grid points and consequently
the center of the microbridge where the order parameter goes to
zero (the phase slip center) is situated between grid points.
Therefore any quantity in the phase slip center (PSC) is in fact
calculated at a distance $\delta x=\xi_0/2 \ll \xi(T)$ from the
PSC.

For the geometrical parameters of the film we used the following
values: $S_2/S_1=10$, $L_2=500 \xi_0$ and $L_1$ was varied from 15
$\xi_0$ (at $T=0.92 T_c$) up to 45 $\xi_0$ (at $T=0.99 T_c$). In
our calculations we used $\tilde{\tau}_{in}=4-1000$ which covers
typical values for many low temperature superconductors (for
example \cite{Watts-Tobin} in Nb $\tilde{\tau}_{in} \simeq 10^2$
and in Al $\tilde{\tau}_{in} \simeq 10^3$ ). In the wide part of
the film we used $\tilde {\tau}_{in}=5$ which allows us to
decrease $L_{in}$ and the length $L_1$ in order to optimize the
calculation time.

To find the current-voltage characteristics of the superconducting
microbridge we applied a large voltage $\pm V_{ext}$ to the normal
reservoirs which induces a large current ($I>I_c$) resulting in a
resistive state in the microbridge. Than we decrease $V_{ext}$ in
a step-wise manner and we find the time averaged difference of the
electrostatic potentials between the ends of the microbridge
($\overline V=\overline {\varphi}(-L/2)-\overline {\varphi}(L/2)$)
as function of $V_{ext}$. In a similar way it is easy to find the
critical voltage $V_{ext}^c$ (or critical current $I_c$) of the
superconducting microbridge when the superconducting state becomes
unstable (by increasing $V_{ext}$ from small values). As a result
we find the dependence $\overline {V}(V_{ext}/V_{ext}^c)$ which
practically coincides with the dependence $\overline {V}(I/I_c)$.
In Fig. 2 we plot the time-dependence of the current in the
microbridge which illustrates that indeed $I$ varies weakly in
time and the constant $V_{ext}$ induces an almost constant $I$ in
our model system even when the microbridge is in the resistive
state.

\section{Results}

\begin{figure}[hbtp]
\includegraphics[width=0.5\textwidth]{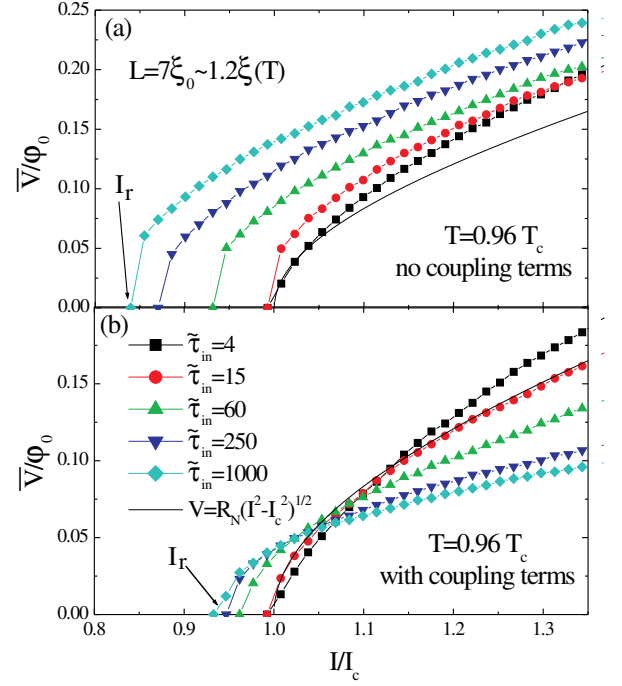}
\caption{(color online). Current-voltage characteristics (in the
regime of decreasing current) for a short microbridge with length
$L=7 \xi_0$ at $T=0.96 T_c$ calculated for different values of the
inelastic relaxation time in the absence (a) and the presence (b)
of coupling terms in Eqs. 1(a,b). Solid curve is the IV
characteristic of a short microbridge in the absence of
nonequilibrium effects (see for example \cite{Aslamazov0}).}
\end{figure}

In Fig. 3 we show the current-voltage characteristics of a short
microbridge with length $L=7 \xi_0$ at $T=0.96 T_c$ calculated for
different values of the inelastic relaxation time $\tau_{in}$ in
the regime of decreasing applied current (the length and
temperature were chosen close to the parameters of the experiment
of Ref. \cite{Octavio} and for tin $\tilde{\tau}_{in}\sim 200$).
Firstly, we should note the strong influence of the coupling terms
in Eqs. 1(a,b) on the IV characteristics: voltage increases
(decreases) with increasing $\tau_{in}$ at $I\gtrsim I_c$ in case
of the absence (presence) of coupling terms. Secondly, in both
cases the IV curves are hyperbolic-like with no 'foot'-like
feature and with hysteresis for relatively large $\tau_{in}$ (the
transition to the superconducting state occurs at the retrapping
current $I_r$ which could be smaller than the critical current
$I_c$ for the transition from the superconducting to the resistive
state).

We plot in Fig. 4 the energy dependence of $\delta f_L$ averaged
over one oscillation period $T_{|\Delta|}$ taken in the phase slip
center. The effect of the coupling terms are clearly visible by
comparing Fig. 4(a) and 4(b). The $\overline{\delta f}_L^{PSC}$ is
on the average negative in case the coupling terms are absent and
positive when they are included.
\begin{figure}[hbtp]
\includegraphics[width=0.45\textwidth]{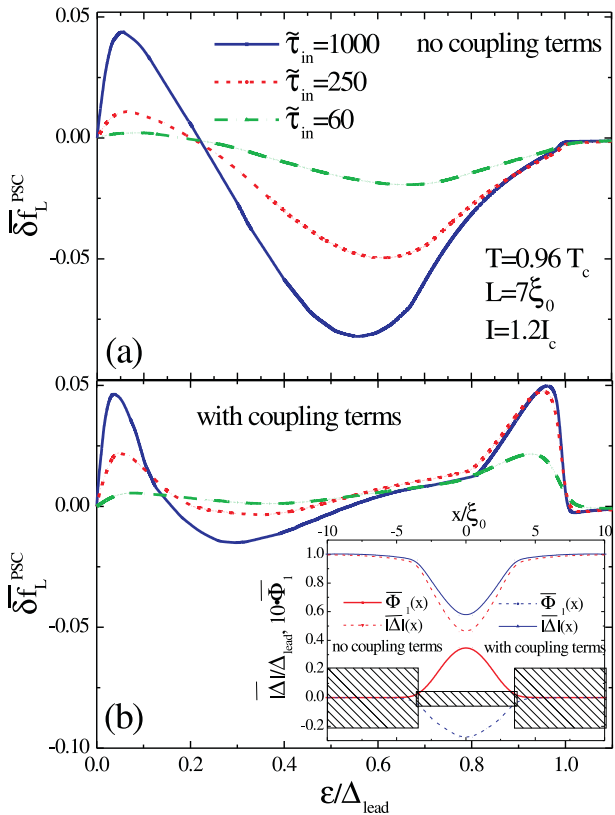}
\caption{(color online). Typical energy dependence of the time
averaged $\delta f_L$ in the center of the short microbridge (i.e.
in the phase slip center) in the absence (a) and the presence (b)
of the coupling terms in Eqs. 1(a,b). In the inset of Fig. 4(b) we
show the distribution of the time averaged $\Phi_1$ and $|\Delta|$
along the microbridge and the leads (schematically shown).}
\end{figure}

Let us now discuss the origin of the sign of $\overline{\delta
f}_L^{PSC}$ for the different energies. First we consider the case
when the coupling terms are neglected. Let us simplify Eq. 1(a) to
\begin{equation}
\frac{\partial \delta f_L }{\partial t }= -\frac{\delta
f_L}{\tau_{in}}-\frac{R_2}{N_1}\frac{\partial f_L^0}{\partial
\epsilon}\frac{\partial |\Delta|}{\partial t},
\end{equation}
where we omit the diffusive and coupling terms. If we average Eq.
(7) over $T_{|\Delta|}$ then we obtain
\begin{equation}
\overline{\delta f}_L=-\tau_{in}\frac{\partial f_L^0}{\partial
\epsilon}\int_{0}^{T_{|\Delta|}}\frac{R_2}{N_1}\frac{\partial
|\Delta|}{\partial t}dt.
\end{equation}
If $R_2$ and $N_1$ are solely a function of $|\Delta|$ then
$\overline{\delta f}_L=0$. But in general the spectral functions
$R_2$ and $N_1$ are a function of two variables $|\Delta|$, $|Q|$
(at fixed x and $\epsilon$ - see Eq. (2)). To get insight how the
ratio $R_2/N_1$ changes with varying $|Q|$ at {\it fixed}
$|\Delta|$ one may solve Eq. (2) with zero second derivative and
find that the ratio $R_2/N_1$ decreases when $\epsilon \gtrsim
|\Delta|$ and increases when $\epsilon \lesssim |\Delta|$ for
large $|Q|$.

In Fig. 5 we present the time dependence of $|\Delta|$, $Q$ and
$\Phi_1$ in the absence of the coupling terms in Eqs. 1(a,b).
First of all we should note that the amplitude of the oscillations
of $|\Delta|$ is smaller than $\Delta_{lead}$. The reason for this
effect is the large characteristic time scale for the variation of
$|\Delta|$ in comparison with the one of $\delta \phi$. Indeed it
is known that nonequilibrium effects may considerably slow down
the dynamics of $|\Delta|$ \cite{Pals,Tinkham2,Vodolazov2}. When
$|\Delta|$ increases (decreases) its value is smaller (larger)
than one could expect from the static dependence
$|\Delta|_{stat}(\delta\phi)$ (see for example Eq. (1.3) in
\cite{Aslamazov0}) for the given value of $\delta \phi(t) \sim
Q(t)L$. For example when $\delta \phi$ reaches zero (which
corresponds to the moment in time when $Q=0$ after the phase slip
event in Fig. 5) the order parameter is still much smaller than
$\Delta_{lead}$.

Due to the time delay in the variation of $|\Delta|$ the
supervelocity $Q\sim \delta \phi/L$ is different for the same
values of $|\Delta|$ taken at different times during the period
$T_{|\Delta|}$ (see dashed lines in Fig. 5). Therefore the ratio
$R_2/N_1$ in the increasing region of $|\Delta|$ is larger
(smaller) for $\epsilon \gtrsim |\Delta|(x,t)$($\epsilon \lesssim
|\Delta|(x,t)$) than in the decreasing region. It results in
$\overline {\delta f}_L<0$ in the phase slip center for $\epsilon
\gtrsim \overline{|\Delta|}^{PSC}$ and $\overline {\delta f}_L>0$
for $\epsilon \lesssim \overline{|\Delta|}^{PSC}$ (see Fig. 4(a)).
The time averaged potential $\Phi_1$ is positive in the
microbridge (see inset in Fig. 4(b)) which implies 'heating' of
quasiparticles and suppression of the superconducting properties.
The larger $\tau_{in}$ the stronger the deviation form equilibrium
(see Eq. (8) and Figs. 4(a)) and the larger the hysteresis (see
Fig. 1(a)).
\begin{figure}[hbtp]
\includegraphics[width=0.5\textwidth]{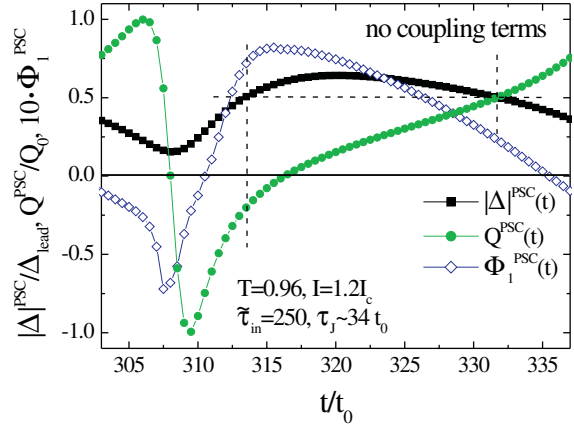}
\caption{(color online). Time dependence of the order parameter
$|\Delta|$, the supervelocity Q and the potential $\Phi_1$ in the
center of the short microbridge (parameters are the same as in
Fig. 4) at $I=1.2I_c$ and $\tilde{\tau}_{in}=250$ as found from a
numerical solution of Eqs. (1-3) without coupling terms. The
horizontal dashed line marks $|\Delta|=\Delta_{lead}/2$ to
illustrate the corresponding different Q values at two moments in
time.}
\end{figure}

Consider now the effect of the coupling terms in Eqs. 1(a,b). The
term $D\nabla (j_{\epsilon}f_T)= 2D\nabla(N_2R_2Q f_T)$ in Eq.
1(a) may be considered as an additional source of nonequilibrium
which is nonzero in the energy interval $\delta \epsilon \simeq
Q^2$ near local $|\Delta|(x)$ (in the spatially uniform case with
$Q=0$ and $\tau_{in}=\infty$ one has $N_2R_2 \sim$ $|\Delta|(x)
\delta(\epsilon-|\Delta|(x)$) - see for example Ref.
\cite{Schmid}). From Eq. 1(b) we may roughly estimate $f_T \sim -
e\varphi \partial f_L^0/\partial {\epsilon}$ and the sign of the
coupling term $D\nabla(j_{\epsilon} f_T)$ is defined mainly by the
product ${\bf Q \cdot E}$ (for discussion of the effect of the
coupling terms see also Ref. \cite{Vodolazov2}).
\begin{figure}[hbtp]
\includegraphics[width=0.5\textwidth]{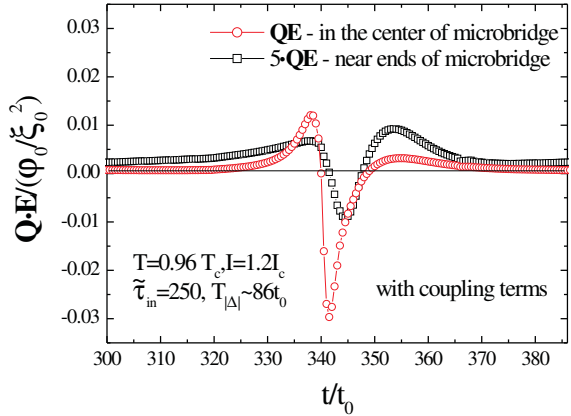}
\caption{(color online). Time dependence of the product ${\bf
Q\cdot E}$ in the phase slip center and near the ends of the short
superconducting microbridge during $T_{|\Delta|}$ as found from a
numerical solution of Eqs. (1-3) including the coupling terms. }
\end{figure}

Our calculations show (see Fig. 6) that near the superconducting
leads the product ${\bf Q \cdot E}$ is mainly (in time) positive
and in the phase slip center it can be both positive and negative
(due to the sign change of Q - see Fig. 5). It results in positive
$\overline{\delta f_{L}}$ at $\epsilon \sim \Delta_{lead}$ - see
Fig. 4(b) (at these energies the ends of the microbridge are
determinative because near the leads $|\Delta|$ has a low
oscillation amplitude and $\overline{{\bf Q \cdot E}}>0$). At
lower energies $\epsilon \sim \overline{|\Delta|}_{PSC} <
\Delta_{lead}$ the main source of nonequilibrium is situated in
the center of the microbridge where both ${\bf Q \cdot E}$ and
$\partial |\Delta|/\partial t$ changes sign during the
oscillations and $\overline{\delta f_L}$ is negative except at low
energies $\epsilon \ll \overline{|\Delta|}^{PSC}$ (see Fig. 4(b)).
The net effect is a negative time averaged $\overline {\Phi}_1$
(see inset in Fig. 4(b)) and therefore an enhancement of the
superconducting properties which explains the decrease of the
voltage at fixed current $I\gtrsim I_c$ with increase of
$\tau_{in}$ (see Fig. 3(b)). But at low voltages the effect of the
coupling terms ($\sim {\bf QE}$) becomes smaller and the role of
'heating' increases. This subsequently leads to an increase of the
voltage at fixed current $I\sim I_c$ with increasing $\tau_{in}$
and to an hysteresis of the IV curves (but which is smaller than
in the case without coupling terms - compare Figs. 5(a) and 5(b)).

\section{Discussion}

The origin of the here predicted time averaged 'heating' of
quasiparticles is different from Joule dissipation. Indeed, the
source of nonequilibrium $\sim \partial |\Delta|/\partial t$
changes sign during oscillations of the order parameter while the
term $I\cdot V$ is always positive. Time averaged 'heating'
appears only due to the dependence of the spectral functions both
on $|\Delta|$ and $Q$ and due to the difference in time variation
of $|\Delta|$ and $\delta \phi$. Energy for 'heating' of the
quasiparticles (and 'cooling' due to the term $\nabla
(j_{\epsilon} f_T)$ in the kinetic equation) comes from the energy
$I\cdot V$ delivered by the external source of the current (this
energy goes also for example to the 'heating' of phonons due to
presence of the relaxation term in the kinetic equations).

Our result is opposite to the one of Refs. \cite{Aslamazov,Schmid}
where a time averaged 'cooling' of quasiparticles and an
enhancement of superconductivity was found at high voltages (when
$T_{|\Delta|}\ll \tau_{in}$) in the absence of the coupling terms
in Eqs. 1(a,b). The differences with our calculations are that we
take into account: i) the backaction of the nonequilibrium
quasiparticle distribution function on the dynamics of $|\Delta|$;
and ii) the dependence of the spectral functions $N_1$, $N_2$ and
$R_2$ on {\it both} $|\Delta|$ and $|Q|$. There is also one more
difference with Refs. \cite{Aslamazov,Schmid} - we did not neglect
the relaxation term in Eq. (7) at high voltages (see Eq. (3) in
Ref. \cite{Aslamazov} and Eq. (16) in Ref. \cite{Schmid}). By
omitting this term one may obtain 'cooling' with $\overline{\delta
f}_L > 0$ if one chooses the initial condition $\delta f_L=0$ when
$|\Delta|=\Delta_{lead}$ in the center of the microbridge (and
when the spectral functions $R_2$ and $N_1$ depend only on
$|\Delta|$ as was assumed in Refs. \cite{Aslamazov,Schmid}). But
this choice of the initial condition is not obvious. By choosing a
different initial condition (large negative $\delta f_L$ at
$|\Delta|=\Delta_{lead}$) one may obtain 'heating' in this case
too. Only the presence of the relaxation term in Eq. (7) resolves
this problem, because it defines undoubtedly the time averaged
value of $\delta f_L$ (see Eq. (8)). Note that due to model
assumptions for spectral functions of Refs.
\cite{Aslamazov,Schmid} $\overline{\delta f}_L$ has to be equal to
zero at any voltage if one does not neglect the relaxation term.
From a physical point of view the relaxation term in Eq. (7) tells
us that the accumulation of the 'heat'/'cold' (after sudden change
of the current for example) occurs on a time scale which is
proportional to $\tau_{in} \gg T_{|\Delta|}$ and one should take
it into account to find the true value of $\overline{\delta f}_L$
and $T_{|\Delta|}$ corresponding to the given value of the
current. It provides a strong dependence of the IV characteristics
and $\overline {\delta f}_L$ on $\tau_{in}$ even at high voltages
(see Figs. 3(a) and 4(a)) while in Refs. \cite{Aslamazov,Schmid}
no dependence on $\tau_{in}$ was found in this limit.

According to our numerical calculations the IV characteristics of
short 'dirty' superconducting microbridges are hyperbolic-like and
hysteretic (at relatively large $\tau_{in}$ and/or low
temperatures) in the temperature interval $0.92<T/T_c<1$
(calculated and not presented here the IV characteristics at
$T/T_c$=0.92-0.99 are qualitatively similar to those of Figs.
3(a,b)). We believe that the experimentally found 'foot'-like
features in the IV characteristics of short superconducting
microbridges is connected with the geometry of the studied
microbridges - they were short and {\it wide} (in experiments
\cite{Guthmann,Gubankov,Octavio} the width $W\simeq 2L$). For such
a geometry the current density distribution over the width of the
microbridge (constriction) should be nonuniform with sharp peaks
near the edges of the microbridge \cite{Aslamazov2}. In Ref.
\cite{Vodolazov} it was shown that a nonuniform current density
distribution in the superconductor may lead to slow vortex motion
at low currents and phase slip lines at high currents
\cite{Vodolazov} and to an IV characteristic (see for example Fig.
4 in Ref. \cite{Vodolazov}) which resembles the experimental
results of Refs. \cite{Guthmann,Gubankov,Octavio}. Moreover the
sharp transition from low to high voltages occurs at $V_c$ which
is inversely proportional to $\tau_{in}$ \cite{Vodolazov} - the
same result was found in the experiment on tin superconducting
microbridges \cite{Octavio}. Therefore, we predict that the IV
characteristics of short ($L \lesssim \xi(T)$) superconducting
microbridges should change from hyperbolic-like (in case when $W
\ll L$) to the one with a well-pronounced 'foot'-like structure at
low voltages when $W\gtrsim L$. Good candidates are Sn, In, Pb and
Al with relatively large coherence length ($\xi(T) \gtrsim 500 nm$
at $T \sim T_c$). We expect that our results will be tested in the
near future because recently it has become possible to fabricate
microbridges with length $L <$ 500 nm and width $W \gtrsim 50$ nm
(see for example Ref. \cite{Wang,Hazra}) which satisfy the
condition $W \ll L \lesssim \xi(T)$ at $0.9<T/T_c<1$.

\begin{acknowledgments}

This work was supported by the Russian Foundation for Basic
Research, Russian Agency of Education under the Federal Target
Programme "Scientific and educational personnel of innovative
Russia in 2009-2013", Flemish Science Foundation (FWO-Vl) and the
Belgian Science Policy (IAP).

\end{acknowledgments}


\begin{references}

\bibitem{Likharev} K. K. Likharev, Rev. Mod. Phys. {\bf 51}, 101 (1979).

\bibitem{Wang} J. Wang, X. Ma, S. Ji, Y. Qi, Y. Fu, A. Jin, L. Lu, C. Gu, X. C. Xie, M. Tian, J. Jia,
and Q. Xue, Nano Res. {\bf 2} 671 (2009).

\bibitem{Hazra} D. Hazra, L. M. A. Pascal, H. Courtois, and A. K. Gupta,
Phys. Rev. B {\bf 82} 184530 (2010).

\bibitem{Aslamazov0} L. G. Aslamazov and A. F. Volkov, in {\it Nonequilibrium
Superconductivity}, edited by D. N. Langenberg and A. I. Larkin
(Elsevier, Amsterdam, 1986), Chap. 2.

\bibitem{Guthmann} G. Guthmann, J. Maurer, M. Belin, J. Bok, and A. Libchaber,
Phys. Rev. B {\bf 11}, 1909 (1975).

\bibitem{Gubankov} V. N. Gubankov, V. P. Koshelets, and G. A.
Ovsyannikov, Zh. Eks. Teor. Fiz. {\bf 73}, 1435 (1977).

\bibitem{Octavio} M. Octavio, W. J. Skocpol, and M. Tinkham, Phys. Rev. B {\bf 17}, 159 (1977).

\bibitem{Aslamazov} L. G. Aslamazov and A. I. Larkin, Zh. Eks. Teor. Fiz. {\bf 70}, 1340 (1976)
[Sov. Phys. JETP {\bf 43}, 698 (1976)].

\bibitem{Schmid} A. Schmid, G. Sch\"{o}n, and M. Tinkham, Phys. Rev. B {\bf 21} 5076 (1980).

\bibitem{Golub}  A. A. Golub, Zh. Eks. Teor. Fiz. {\bf 71}, 341
(1976)[Sov. Phys. JETP {\bf 44}, 178 (1976)].

\bibitem{Tinkham} M. Tinkham, {\it Introduction to superconductivity}, (McGraw-Hill, NY, 1996), p. 417.

\bibitem{Schmid2} A. Schmid and G. Sch\"{o}n, J. Low Temp. Phys. {\bf
20}, 207 (1975).

\bibitem{Larkin} A.I. Larkin and Yu. N. Ovchinnikov, Zh. Eksp. Teor. Fiz. {\bf 73}, 299 (1977)
[Sov. Phys. JETP {\bf 73}, 155 (1977)].

\bibitem{Kramer} L. Kramer and R.J. Watts-Tobin, Phys. Rev. Lett.
{\bf 40}, 1041 (1978).

\bibitem{Watts-Tobin} R.J. Watts-Tobin, Y. Kr\"ahenb\"uhl, and L. Kramer,
 J. Low Temp. Phys. {\bf 42}, 459 (1981).

\bibitem{Keizer} R. S. Keizer, M. G. Flokstra, J. Aarts, and T. M. Klapwijk, Phys. Rev. Lett. {\bf 96}, 147002 (2006).

\bibitem{Pals} J. A. Pals and J. Wolter, Phys. Lett. {\bf 70 A}, 150 (1979) (see also Ref. \cite{Tinkham}, p.
414).

\bibitem{Tinkham2} M. Tinkham, in {\it Non-Equilibrium Superconductivity,
Phonons and Kapitza Boundaries}, edited by K. E. Gray (Plenum, New
York, 1981).

\bibitem{Vodolazov2} D. Yu. Vodolazov and F. M. Peeters, Phys. Rev. B {\bf 81}, 184521 (2010).

\bibitem{Aslamazov2} L.G. Aslamazov and A. I. Larkin, Zh. Eks. Teor. Fiz. {\bf 68}, 766 (1975)[Sov. Phys. JETP {\bf 41}, 381 (1975)].

\bibitem{Vodolazov} D. Yu. Vodolazov and F. M. Peeters, Phys. Rev. B {\bf 76} 014521 (2007).

\end{references}
\end{document}